\documentclass[epj]{svjour}
\usepackage{graphics}
\begin{document}
\title{Interplay of Rashba spin orbit coupling and disorder in the
  conductance properties of a four terminal junction device}
\author{Sudin Ganguly\thanks{\email{sudin@iitg.ernet.in}} \and Saurabh Basu\thanks{\email{saurabh@iitg.ernet.in}}
}                     % Do not remove

\institute{Indian Institute of Technology Guwahati, Guwahati,
  Assam-781039}
\date{Received: date / Revised version: date}
\abstract{
We report a thorough theoretical investigation on the quantum
transport of a disordered four terminal device in the presence of
Rashba spin orbit coupling (RSOC) in two dimensions. Specifically we
compute the behaviour of the longitudinal (charge) conductance, spin
Hall conductance and spin Hall conductance fluctuation as a function
of the strength of disorder and Rashba spin orbit interaction using
the Landauer B${\ddot\mathrm{u}}$ttiker formalism via Green's function
technique. Our numerical calculations reveal that both the
conductances diminish with disorder. At smaller values of the RSOC
parameter, the longitudinal and spin Hall conductances increase, while
both vanish in the strong RSOC limit. The spin current is more
drastically affected by both disorder and RSOC than its charge
counterpart. The spin Hall conductance fluctuation does not show any
universality in terms of its value and it depends on both disorder as
well as on the RSOC strength. Thus the spin Hall conductance
fluctuation has a distinct character compared to the fluctuation in
the longitudinal conductance. Further one parameter scaling theory is
studied to assess the transition to a metallic regime as claimed in
literature and we find no confirmation about the emergence of a
metallic state induced by RSOC.
\PACS{
      {PACS-key}{discribing text of that key}   \and
      {PACS-key}{discribing text of that key}
     } % end of PACS codes
} %end of abstract
\titlerunning{Interplay of Rashba SOC and disorder}
\authorrunning{S. Ganguly \and S. Basu}

\maketitle
\section{Introduction}
\label{intro}
Spintronics or spin based electronics focuses on transportation of
electron spins in a variety of semiconducting materials
\cite{wolf,dd,murakami}. The prospects of obtaining a dissipation less
spin current along with a large number of applications possible, have
generated intense activities from both the theoreticians and the
experimentalists.

While there were early attempts of injecting spin polarized charge
carries (from ferromagnetic metals) in non magnetic semiconductors
(thereby converting them to dilute magnetic semiconductors) which have yielded very limited success. However the recent discovery of intrinsic spin
Hall effect (SHE) in p-type semiconductors, that originate from an
effective magnetic field and induced by a Berry phase makes the
up and down spin electrons drift in opposite directions contribute immensely to the ongoing experimental work on spin manipulation phenomena in real materials \cite{murakami}. In two dimensional electron gases
(2DEG), one can hope to realize some of the interesting physics in this regard in presence of Rashba spin orbit coupling (RSOC) \cite
{sinora}.

In this work we shall concentrate on Hall effect (for both charge and
spin) induced by RSOC in a four terminal junction device in presence of
disorder. The existing work on the subject of interplay of disorder
and RSOC on the conductance properties of such junction devices is
limited and not beyond controversy \cite {shen,burkov,schhemann}.

In a simple language, the existence of RSOC distinguishes the up and
down spin electrons and hence a potential gradient applied in the
$x$-direction, the opposite spins drift towards mutually opposite
$y$-direction and an accumulation of spins can be observed along the
edges of the sample, thereby yielding SHE. In experiments such
accumulation of spins could be optically detected by using the Kerr
rotation spectroscopy \cite{kato,sih}. Now, as disorder is mostly an
inextricable component in real systems, we seek to study the effect of
disorder on the spin Hall conductance properties of disordered
samples. The issue is particularly relevant because of a number of
reasons. On one hand, there are studies that claim that SHE can be
strongly suppressed by disorder effects \cite{inoue} signaling onset
of spin relaxation and diffusion \cite{araki}, while the other school
of thought says that since the spins are tightly locked with their
momenta (in presence of RSOC), SHE should be resilient to a disorder
potential \cite{sante}. Also there are suggestions of emergence of a
metallic behaviour at a certain critical disorder strength in presence
of RSOC, which however vanishes (as it should be) in the vanishing of
the Rashba coupling \cite{shen}. Therefore the scenario warrants a
more complete study to settle some of the crucial issues in these
regards.

Also, the universal charge conductance fluctuation \cite{lee-stone} is
one of the most striking quantum transport feature in mesoscopic
physics, which is nothing but the consequence of quantum interference
when the inelastic diffusion length exceeds sample dimensions. It is
well known that the conductance fluctuations of the order of $e^2/h$
are universal features of quantum transport in the low temperature
limit. In particular, they are independent of both the degree of
disorder and the sample size at zero temperature. Analogously, one can
study the fluctuations in spin Hall conductance and it is found that,
there exists a universal spin Hall conductance fluctuations, which has
a finite value, independent of other system details. However, this
fluctuation is a function of the spin orbit interaction strength
\cite{qiao,qiao1}. In this paper we study the variation of spin Hall
conductance fluctuation as a function of the disorder strength and
strength of RSOC, in order to observe the universality and independent
nature of spin Hall conductance fluctuation.

Motivated by the preceding discussion, we aim to investigate in a four
terminal junction device the spin Hall conductance, the usual charge
conductance (referred to as the longitudinal conductance), the
universal spin Hall conductance fluctuation (the root mean square
value of which $\left(=0.36\left(\frac{e}{8\pi}\right)\right)$ is
independent of other parameters of the Hamiltonian and is thus
attributed as `universal' \cite{qiao}.

Further we have critically examined the onset of a metallic regime at
a certain critical disorder strength for a given value of the spin
orbit coupling parameters as discussed by Sheng \textit{et. al}
\cite{shen} form the behaviour of the longitudinal conductance. Their
claim is as the spin orbit coupling vanishes, the critical disorder
strength too vanishes, thereby yielding an absence of a metal
insulator transition in two dimensions, a result that is well known.

While we are thoughtful about the applicability of a one parameter
scaling theory \cite{gang_of_4} in presence of RSOC, however went
ahead assuming it is at least appreciable for a particular value of
the spin orbit coupling parameter. We do not find any convincing
evidence in a Rashba coupled disordered four terminal junction or at
best the results do not help in arriving at unambiguous inferences
about it. Our inference on the `possible' absence of an emerging
metallic phase is justified as the Rashba term preserves the time
reversal symmetry, an essential ingredient for the current `echo'
associated with the quantum interference effects that give rise to
weak localization. An external magnetic field would have radically
modified theses arguments.

In the scaling theory, the logarithmic
derivative of conductance with respect to the system size is defined
by a single quantity, usually denoted by $\beta$, which depends only
on the conductance itself and not individually on energy, system size
and disorder. Weak localization in two dimensional systems can be best
characterized by the relation \cite{gang_of_4},
\begin{equation}
\beta = \frac{d \ln{G_L}}{d\ln{L}}
\end{equation}
where, $G_L$ is the charge conductance and $L$ is the length of the
conductor. Therefore the scenario warrants a more complete study to
settle some of the crucial issues in these regards.

We organize our paper as follows. The theoretical formalism leading to
the expressions for the spin Hall and longitudinal conductances using
Landauer B${\ddot\mathrm{u}}$ttiker formula are presented in section
II. Section III includes an elaborate discussion on the results
obtained for the longitudinal conductance, spin Hall conductnace and
fluctuation in the spin Hall conductance that are helpful in depicting
the interplay of disorder strength and RSOC parameter in details.

\section{Theoretical formulation}

\subsection{System and the Hamiltonian}

\begin{figure}
\centering
\resizebox{0.4\textwidth}{!}{%
  \includegraphics{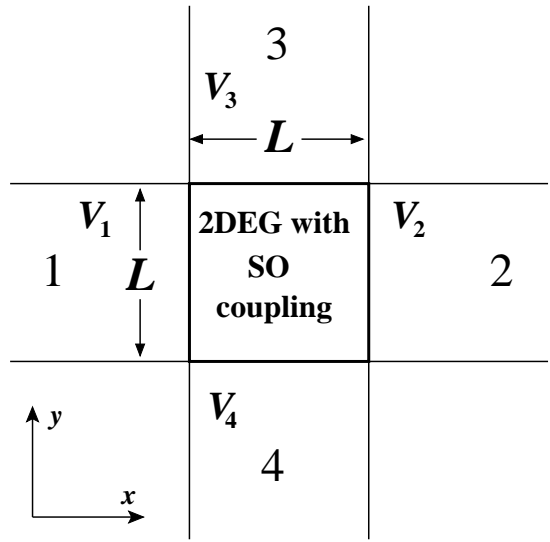}
}
\caption{Cross-shaped device with four semi-infinite metallic
  leads. The spin-orbit coupling exists in the central region only,
  and the effect of the semi-infinite leads is treated exactly through
  the self-energy terms.}
\label{setup}
\end{figure}

In order to observe the spin Hall effect, we choose a four-probe
measuring set-up as shown in Fig.\ref{setup}. Here the four ideal
(disorder free) semi-infinite leads are attached to the central
conducting region which in our case is a square lattice with Rashba
spin orbit interaction. An unpolarized charge current is allowed to
pass through the longitudinal leads along lead-1 and lead-2 (see
Fig.\ref{setup}) inducing spin Hall current in the transverse
directions that is along lead-3 and lead-4. The discrete spectrum of
the longitudinal region is assumed to be described within the
framework of tight-binding approximation assuming only
nearest-neighbor hopping. The Hamiltonian for the entire system can be
represented by a sum of three terms, which are,

\begin{equation}
H=H_{sq}+H_{leads}+H_{coupling}
\label{h1}
\end{equation} 

\noindent The first term represents the Hamiltonian for the system
with a square lattice geometry with Rashba spin orbit interaction and
is defined by,

\begin{eqnarray}
H_{sq}=\sum\limits_{i,\sigma} \epsilon_i c_{i\sigma}^{\dagger}
c_{i\sigma} + t\sum\limits_{\langle ij\rangle,\sigma}
c_{i\sigma}^{\dagger} c_{j\sigma} + \nonumber\\\alpha\sum\limits_{i}
\left[\left(c_{i\uparrow}^{\dagger} c_{i+\delta_x \downarrow} -
  c_{i\downarrow}^{\dagger} c_{i+\delta_x \uparrow}\right)
  \right.\nonumber \\ \left.  -i \left(c_{i\uparrow}^{\dagger}
  c_{i+\delta_y \downarrow} + c_{i\downarrow}^{\dagger} c_{i+\delta_y
    \uparrow}\right) \right]
\label{h2}
\end{eqnarray} 
Here, $\epsilon_i$ is the random on-site potential energy chosen from
a uniform rectangular distribution ($-W$ to $W$), $c_{i}^{\dagger}$
and $c_{i}$ correspond to the creation and annihilation operators
respectively, for an electron at the site $i$ of the conductor. Here
$t=\frac{\hbar}{2m^*a_0^2}$ ($m^*$: effective mass and $a_0$: lattice
constant) is the hopping integral, $\alpha$ is the Rashba coupling
strength. $\delta_{x,y}$ are the unit vectors along $x,y$ direction.

The four metallic leads attached to the conductor are considered to be
semi-infinite and ideal. The leads are described by a similar
non-interacting single particle Hamiltonian as written below,

\begin{equation}
H_{leads}=\sum\limits_{i=1}^4 H_{i},
\label{h3}
\end{equation} 
where

\begin{equation}
H_{i}=\epsilon_{L}\sum\limits_{n} c_{n}^{\dagger} c_{n} +
t_{L}\sum\limits_{\langle mn\rangle} c_{m}^{\dagger} c_{n}
\label{h4}
\end{equation} 
Similarly, the conductor-to-lead coupling is described by the
following Hamiltonian.
\begin{equation}
H_{coupling}=\sum\limits_{i=1}^4 H_{coupling,i}\quad ,
\label{h5}
\end{equation} 
where
\begin{equation}
H_{coupling,i}=t_{C}\left[c_{i}^{\dagger}c_{m}+c_{m}^{\dagger}c_{i}\right]
\label{h6}
\end{equation} 
In the above expression, $\epsilon_L$ and $t_{L}$ stand for the site
energy and nearest-neighbor hopping between the sites of the
leads. The coupling between the conductor and the leads is defined by
the hopping integral $t_{C}$. In Eq. \ref{h6}, $i$ and $m$ belong to
the boundary sites of the square lattice and leads, respectively. The
summation over $i$ accounts for the four attached leads.

\subsection{Formulation of longitudinal and spin Hall conductances}

For the four-probe case, where pure spin current is expected to flow
through the transverse leads, due to the flow of charge current
through the longitudinal leads, the longitudinal and spin Hall
conductances are defined as \cite{mou},

\begin{equation}
G_L = \frac{I_2^q}{V_1 - V_2} 
\label{def_gl}
\end{equation}
and,
\begin{equation}
G_{SH} = \frac{\hbar}{2e}\frac{I_3^s}{V_1-V_3} =
\frac{\hbar}{2e}\frac{I_3^\uparrow - I_3^\downarrow}{V_1-V_3}
\label{def_gs}
\end{equation}
where $I_2^q$ and $I_3^s$ are the charge and spin current flowing
through the lead-2 and lead-3 respectively. $V_m$ is the potential at
the $m$-th lead. $I_3^\uparrow$ and $I_3^\downarrow$ are the up and
down spin currents flowing in lead-3.

The calculation of electric and spin currents is based on the
Landauer-B${\ddot\mathrm{u}}$ttiker multi-probe formalism
\cite{Buttiker}. The charge and spin currents flowing through the lead
$m$ with potential $V_m$, can be written in terms of spin resolved
transmission probability as \cite {pareek}
\begin{equation}
I_m^q= \frac{e^2}{h} \sum\limits_{n\neq m,\sigma,\sigma^\prime}
\left(T_{nm}^{\sigma\sigma^\prime}V_m
-T_{mn}^{\sigma^\prime\sigma}V_n\right)
\label{iq}
\end{equation}

\begin{equation}
I_m^s= \frac{e^2}{h} \sum\limits_{n\neq m,\sigma^\prime}
\left[\left(T_{nm}^{\sigma^\prime\sigma} -
  T_{nm}^{\sigma^\prime-\sigma}\right)V_m +
  \left(T_{mn}^{-\sigma\sigma^\prime} -
  T_{mn}^{\sigma\sigma^\prime}\right)V_n\right]
\label{is}
\end{equation}
The spin current through lead $m$ can be written in a more compact form, such as,

\begin{equation}
I_m^s = \frac{e^2}{h}\sum\limits_{n\neq m} \left[T_{nm}^{out} V_m - T_{mn}^{in} V_n\right]
\label{in_out}
\end{equation}
where, we have defined two useful quantities as follows,
\begin{eqnarray}
T_{pq}^{in} &=& T_{pq}^{\uparrow\uparrow} +
T_{pq}^{\uparrow\downarrow} - T_{pq}^{\downarrow\uparrow} -
T_{pq}^{\downarrow\downarrow}\\ T_{pq}^{out} &=&
T_{pq}^{\uparrow\uparrow} + T_{pq}^{\downarrow\uparrow} -
T_{pq}^{\uparrow\downarrow} - T_{pq}^{\downarrow\downarrow}
\end{eqnarray}
Physically, the term $\frac{e^2}{h}\sum\limits_{n\neq m} T_{nm}^{out}
V_m$ in Eq. \ref{in_out} is the total spin current flowing from the
$m$-th lead with potential $V_m$ to all other $n$ leads, while the
term $\frac{e^2}{h}\sum\limits_{n\neq m} T_{mn}^{in} V_n$ is the total
spin current flowing into the $m$-th lead from all other $n$ leads
having potential $V_n$.

\noindent The zero temperature conductance,
$G_{pq}^{\sigma\sigma^\prime}$ that describes the spin resolved
transport measurements, is related to the spin resolved transmission
coefficient as \cite {land_cond,land_cond2},
\begin{equation}
G_{pq}^{\sigma\sigma^\prime} = \frac{e^2}{h} T_{pq}^{\sigma\sigma^\prime}(E)
\end{equation}
The transmission coefficient can be calculated as \cite{caroli,Fisher-Lee},

\begin{equation}
T_{pq}^{\sigma\sigma^\prime} = \rm{Tr}\left[\Gamma_p^\sigma G_R
  \Gamma_q^{\sigma^\prime} G_A\right]
\end{equation}
$\Gamma_p^\sigma$'s are the coupling matrices representing the
coupling between the central region and the leads, and they are
defined by the relation \cite{dutta},
\begin{equation}
\Gamma_{p}^{\sigma} = i\left[\Sigma_p^\sigma -
  (\Sigma_p^\sigma)^\dagger\right]
\end{equation}
Here $\Sigma_p^\sigma$ is the retarded self-energy associated with the
lead $p$. The self-energy contribution is computed by modeling each
terminal as a semi-infinite perfect wire \cite{nico}.

The retarded Green's function, $G_R$ is computed as
\begin{equation}
G_R = \left(E - H - \sum\limits_{p=1}^4 \Sigma_p\right)^{-1}
\end{equation}
where $E$ is the electron energy and $H$ is the model Hamiltonian for
the central conducting region. The advanced Green's function is, $G_A
= G_R^\dagger$.

Now, following the spin Hall phenomenology, in our set-up since the
transverse leads are voltage probes, the net charge currents through
lead-3 and lead-4 is zero that is, $I_3^q = I_4^q = 0$. On the other
hand, as the currents in various leads depend only on voltage
differences among them, we can set one of the voltages to zero without
any loss of generality. Here we set $V_2 = 0$. Finally, if we assume
that the leads are connected to a geometrically symmetric ordered
bridge, so, $\frac{V_3}{V_1} = \frac{V_4}{V_1} = \frac{1}{2}$. Now from
Eq.(\ref{in_out}) and Eq.(\ref{def_gs}) we can write the expression of
spin Hall conductance as,

\begin{equation}
 G_{SH} = \frac{e}{8\pi}\left(T_{13}^{out} + T_{43}^{out} +
 T_{23}^{out} - T_{34}^{in} - T_{31}^{in}\right)
\end{equation}
Similarly, from Eq. \ref{def_gl} and Eq. \ref{iq}, the expression of
longitudinal conductance can be written as,
\begin{equation}
G_L = \frac{e^2}{h}\left(T_{21} + \frac{1}{2}T_{32} + \frac{1}{2}T_{42}\right)
\end{equation}
We define the spin Hall conductance fluctuation as,
\begin{equation}
\Delta G_{SH} = \sqrt{\langle G_{SH}^2\rangle - \langle G_{SH}\rangle^2}
\end{equation}
where $\langle ... \rangle$ denotes averaging over an ensemble of
samples with different configurations with disorder strength $W$.

\section{Results and discussion}

We have investigated the interplay of disorder $(W)$ and ROSC
$(\alpha)$ on experimentally measurable quantities such as
longitudinal conductance ($G_L$), spin Hall conductance ($G_{SH}$) and
spin Hall conductance fluctuation ($\Delta G_{SH}$). The results are
likely to be relevant for systems with spin orbit interactions and
since disorder is an indispensable ingredient in crystal lattices, a
competition between $W$ and $\alpha$ will reveal whether they help or
hinder each other with regard to the conductance properties of
junction devices. An important question in this regard is whether
there is any shift in the localization phenomenon in two dimensions
induced by the RSOC \cite{shen}. More concretely, is there a
transition to a metallic state at some critical value of the strength
of RSOC?

Before we start computing the physical quantities, we briefly describe
the values of different parameters used in our calculation. Throughout
our work, we have considered lattice constant, $a = 1$, onsite term,
$\epsilon =\epsilon_L = 0$, hopping term, $t=t_L=t_C=1$. All the
energies are measured in unit of $t$. Further we choose a unit where
$c=h=e=1$. The longitudinal conductance, $G_L$ is measured in unit of
$\frac{e^2}{h}$. The spin Hall conductance, $G_{SH}$ is measured in
unit of $\frac{e}{8\pi}$. The random onsite disorder is modeled by a
rectangular distribution of the form,
\begin{equation}
P\left(\epsilon_i\right)=\frac{1}{2W} \quad \rm{where,}
\quad-W\le\epsilon_i\le W
\label{dis_distribtn}
\end{equation}
All the results obtained below are averaged over 10000 disorder
configurations.

To remind ourselves we have compared all of the above quantities at a
fixed value of energy (or the biasing voltage), namely $E = -2t$. Any
other energy value would yield qualitatively similar results,
excepting at $E = 0$ where $G_{SH}$ identically vanishes \cite{moca,li_hu}.

There is another small point that deserves a mention. We have
investigated $G_L$, $G_{SH}$ and $\Delta G_{SH}$ both over a small
range of the RSOC, that is with $[0:1]$ and also over a much broader
range, namely up to $\alpha = 8$. Of course, the former is a subset of
the latter, but the behaviour of the physical quantities under
consideration show qualitatively different behaviour in the two
regimes and thus warrant distinct discussion.

\subsection{Longitudinal conductance}

In a tight binding model for a two dimensional square lattice, the
energy band width ($BW$) is $8$ ($-4$ to $4$ in units of $t$). Here,
we set range for the disorder strength, as in the interval $\left[0,
  \frac{BW}{2}\right]$ that is $[0,4]$, while for the RSOC strength,
$\alpha$ to be in $\left[0, BW\right]$ or $\left[0,8\right]$.

\begin{figure}
\centering
\resizebox{0.35\textwidth}{!}{%
  \includegraphics{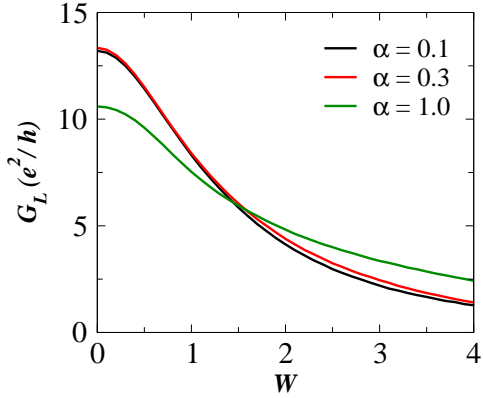}
}
\caption{(Colour online) The longitudinal conductance, $G_L$ ( in
  units of $e^2/h$) is plotted as a function of disorder strength $W$
  (in units of $t$) for different RSOC strengths, $\alpha$. Here we
  consider lower values of $\alpha$, namely $\alpha$ in the region
  $[0:1]$. The plots cross at $W\approx1.5$. All data in this figure
  and all subsequent figures averaged over 10000 disorder
  configuration.}
\label{glvsw}
\end{figure}

The variation of $G_L$ as a function of $W$ for different $\alpha$ is
shown in Fig.\ref{glvsw}. In this figure we have considered for
$\alpha =$ 0.1, 0.3 and 1.0. As expected, $G_L$ falls off with
disorder. However the fall off at lower values of $\alpha$
(e.g. $\alpha$ = 0.1 and 0.3) at large disorder is more than the
corresponding values at large $\alpha$ (e.g. $\alpha =1$). The trend
was reverse at lower values of disorder where there is a crossover at
$W\approx 1.5$. For example, $G_L = 2.5$ (in units of $\frac{e^2}{h}$)
at $W\approx\frac{BW}{2}(=4)$ for $\alpha = 1$ while the corresponding
value is $\approx 1.2$ for $\alpha = 0.1$.

The inference that can be drawn is larger RSOC aides in enhancing the
conductance values at larger disorder. However whether this
enhancement has anything got to do with a transition to a conducting
phase is yet to be seen.

\begin{figure}
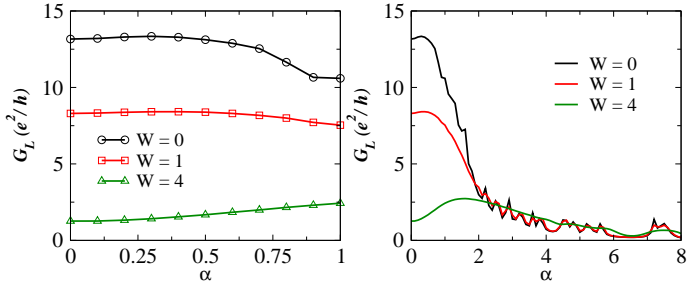

\resizebox{0.5\textwidth}{!}{%
  \includegraphics{fig3a.eps}\includegraphics{fig3b.eps}
}
\caption{(Colour online) (a) $G_L$ is plotted as a function of RSOC
  strength $\alpha$ (in units of $t$) for different disorder strengths
  $(W)$. Here $\alpha$ is varied between $[0:1]$. $G_L$ has weak
  dependence at lower values of $\alpha$. (b) $G_L$ is plotted as a
  function of RSOC strength $\alpha$ for different disorder strengths
  $(W)$. Here we take the range of $\alpha$ as $[0 : BW]$ $(BW =
  8)$. $G_L$ decreases with increasing the strength of RSOC.}
\label{glvsa}
\end{figure}

The variation of $G_L$ as a function of RSOC strength, $\alpha$ for
different disorder strengths are shown in Fig.\ref{glvsa}. In
Fig.\ref{glvsa}(a), $\alpha$ is varied over a small range $[0:1]$. We
observe that in this regime, $G_L$ is not strongly dependent on
$\alpha$ for the values of disorder that we have considered, namely
$W=$ 1 and 4. The disorder free case is included for comparison.

However Fig.\ref{glvsa}(b) shows $G_L$ being plotted for a much wider
rage of $\alpha$, namely $\alpha=$ 0 to $BW(=8)$. In this figure, the
longitudinal conductance decreases with increasing $\alpha$ and at
higher values of $\alpha$, $G_L$ becomes vanishingly small. Thus in
this regime $\alpha$ destroys longitudinal conductance just as the
disorder does. Physically, this means that large values of $\alpha$
denote strong correlated hopping anisotropies, where the hopping
strengths are all different along $\pm x$ and $\pm y$ directions (see
Eq.(\ref{h2})). This emulates disorder effects for the charge carriers
where they see a different environment with regard to hopping.

\subsection{Spin Hall conductance}
\begin{figure}
\centering
\resizebox{0.35\textwidth}{!}{%
  \includegraphics{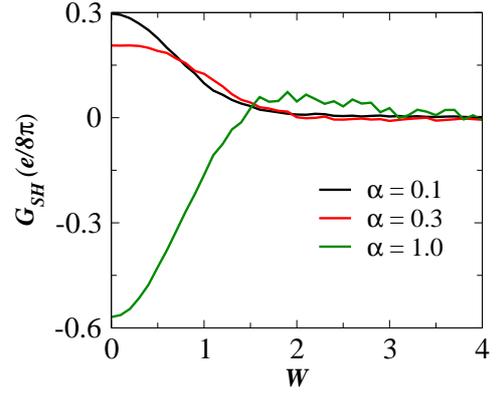}
}
\caption{(Colour online) The spin Hall conductance, $G_{SH}$ (in units
  of $e/8\pi$) is plotted as a function of disorder strength, $W$ for
  different RSOC strengths, $\alpha$. $G_{SH}$ decreases with
  increasing the disorder strength, $W$.}
\label{gsvsw}
\end{figure}

We study the spin Hall conductance, $G_{SH}$ as defined by
Eq.(\ref{def_gs}) in presence of disorder, $W$ and spin orbit
coupling, $\alpha$. Fig. \ref{gsvsw} shows the variation of $G_{SH}$
as a function of disorder strength, $W$ for different values of
$\alpha$. Overall, $G_{SH}$ decreases with increasing disorder
strength. For $\alpha =$ 0.1 and 0.3, $G_{SH}$ starts from positive
values and subsequently vanishes as we increase $W$. While for $\alpha
= 1.0$, $G_{SH}$ starts from a negative value and vanishes at large
$W$. There is a difference with the corresponding data for the
longitudinal conductance, $G_L$ which remains finite at large
disorder.

\begin{figure}
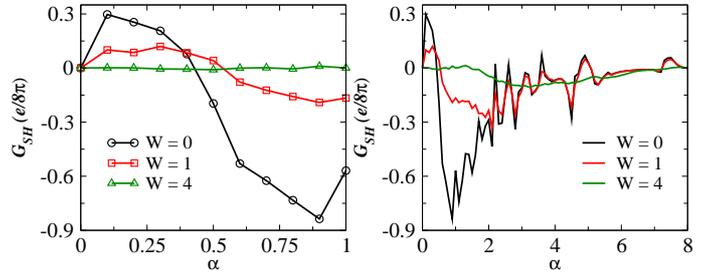

\resizebox{0.5\textwidth}{!}{%
  \includegraphics{fig5a.eps} \includegraphics{fig5b.eps}
}
\caption{(Colour online) Spin Hall conductance, $G_{SH}$ is plotted as
  a function of RSOC strength $\alpha$ for $W =$ 0, 1 and 4. (a)
  $\alpha$ is varying from $[0:1]$. (b) $\alpha$ is varying for a much
  wider range, that is between $[0:8]$.}
\label{gsvsa}
\end{figure}

Fig.\ref{gsvsa} shows the variation of $G_{SH}$ as a function of
$\alpha$ for different values of $W$. In Fig.\ref{gsvsa}(a), we have
plotted $G_{SH}$ for a small range of $\alpha$, namely between
$[0:1]$. As expected, for the pure case ($W = 0$), $G_{SH}$ has a
maximum value (in magnitude). As disorder is introduced, the spin Hall
conductance decreases. In Fig.\ref{gsvsa}(b), $G_{SH}$ is plotted for
a wider range of $\alpha$. It is in tune with the results for
longitudinal conductance that at higher values of $\alpha$, $G_{SH}$
becomes almost zero. However $G_{SH}$ seem to be strongly affected by
disorder. At $W=4$, for all values of $\alpha$, $G_{SH} \approx
0$. There is a fluctuation in $G_{SH}$, the behaviour of which may be
due to the finite size effects \cite{moca}.

\subsection{Spin Hall conductance fluctuation}
As said earlier, the spin Hall conductance fluctuation attains a
finite constant value which is independent of material properties. It
is thus of importance to see how this otherwise constant value
responds to disorder and RSOC.

\begin{figure}
\centering
\resizebox{0.35\textwidth}{!}{%
  \includegraphics{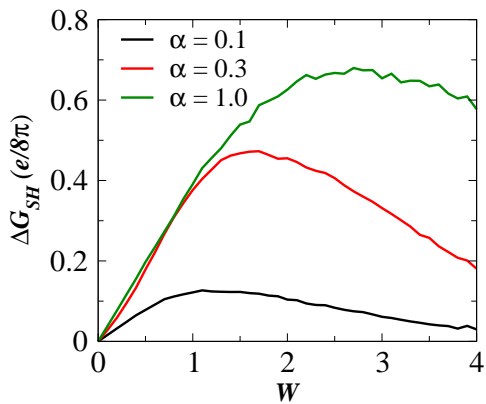}
}
\caption{(Colour online) (a) $\Delta G_{SH}$ (in units of $e/8\pi$) is
  plotted as a function of disorder strength $W$ for different RSOC
  strengths $(\alpha)$. $\Delta G_{SH}$ shows a non monotonic
  behaviour with disorder, $W$.}
\label{gvarvsw}
\end{figure}

In Fig. \ref{gvarvsw}(a), $\Delta G_{SH}$ (in unit of
$\frac{e}{8\pi}$) is plotted as a function of $W$ for different values
of $\alpha$. As we increase $\alpha$ the fluctuation also
increases. For each value of $\alpha$ and at lower values of $W$, it
may be noted that, $\Delta G_{SH}$ increases linearly, and hence it
reaches a maximum value. Beyond this, $\Delta G_{SH}$ decreases with
increasing $W$, yielding a non-monotonic behaviour \cite{qiao}.

\begin{figure}
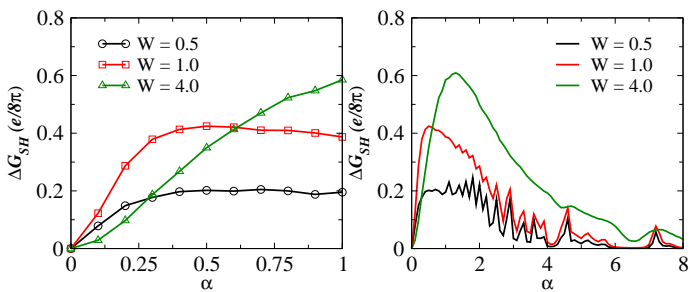

\resizebox{0.5\textwidth}{!}{%
  \includegraphics{fig7a.eps}\includegraphics{fig7b.eps}
}
\caption{(Colour online) (a) $\Delta G_{SH}$ is plotted as a function
  of RSOC strength $\alpha$ for $W =$ 0.5, 1 and 4 for a small range
  of $\alpha$, that is between $[0:1]$. (b) $\Delta G_{SH}$ is plotted
  as a function of RSOC strength $\alpha$ for a much wider range,
  namely between $[0:8]$.}
\label{gvarvsa}
\end{figure}

Fig.\ref{gvarvsa} shows the variation of $\Delta G_{SH}$ as a function
of $\alpha$ for different values of the disorder strength. In
Fig.\ref{gvarvsa}(a), for small disorder values, that is $W=0.5$ and
$1$, $\Delta G_{SH}$ tends to saturate at values, namely 0.2 and 0.4
respectively. As we increase disorder, the fluctuation also increases
and the corresponding plot shows no saturation upto $W=4$. In
Fig.\ref{gvarvsa}(b), we have shown the variation of $\Delta G_{SH}$
for a wider range of $\alpha$, which shows a non-monotonic behaviour,
beyond $\alpha\approx 1$ and ultimately vanishing at large values of
$\alpha$, albeit not without fluctuation. Further the plots for lower
values of disorder (e.g. $W=$ 0.5 and 1, as compared to $W=1$) show
suppressed fluctuations.

It may be noted that in the fluctuation in spin Hall conductance we
did not find any universality in presence of disorder and
RSOC. $\Delta G_{SH}$ depends on the disorder strength, $W$ and the
strength of the RSOC, $\alpha$.

\subsection{One parameter scaling theory}
\begin{figure}
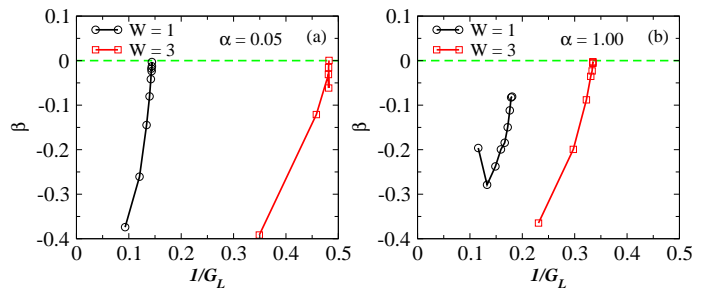

\resizebox{0.5\textwidth}{!}{%
  \includegraphics{fig8a.eps}\includegraphics{fig8b.eps}
}
\caption{(Colour online) $\beta$ is plotted as a function of
  $1/G_L$. (a) $\alpha=0.05$ , (b) $\alpha=1.0$. In both the figures,
  the strengths of disorder, $W$ are 1 and 3. The plots do not cross
  the dashed line $(\beta=0)$ and always remain negative.}
\label{betafig}
\end{figure}

It may be noted that the following observations were made in
Fig.\ref{glvsa}, that the longitudinal conductance, $G_L$ is seen to
increase at larger disorder $\left(W \sim \frac{B}{2}\right)$ as
$\alpha$ increases. At $\alpha=1$, $G_L$ is greater than that
corresponding to other lower values of $\alpha$. This enhancement in
conductance whether signals a transition to a metallic phase as hinted
in Ref.\cite{shen} is to be assessed. To arrive at a conclusion we
perform a one parameter scaling theory \cite{gang_of_4}. For that, we
study the variation of $\beta$ as a function of $1/G_L$ in presence of
RSOC and random onsite disorder. The results are shown in
Fig.\ref{betafig}. In Fig. \ref{betafig}(a), $\beta$ is plotted as a
function of $1/G_L$ for two disorder strengths, namely $W=$ 1 and 3,
where we have kept $\alpha$ to be constant at a very low value, such
as 0.05. In Fig. \ref{betafig}(b), we take a larger value of
$\alpha(=1.0)$. In both Fig.\ref{betafig}, the maximum values of
$\beta$ attain values close to zero, however there is no crossing of
the dashed line $(\beta = 0)$ in Fig.\ref{betafig}. $\beta$ becoming
positive (convincingly staying above the dashed line) would have
indicated a transition to a metallic phase as the conductance directly
scaling with system size is a typical signature for the presence of
extends states and hence metallic behaviour. Hence we can not say
anything about the existence of onset of a metallic regime at a
certain critical disorder strength for a given value of the spin orbit
coupling parameters as informed by Sheng et. al. \cite{shen}.

\section{Summary and Conclusion}
In summary, in the present work we have studied the interplay of
random onsite disorder, $W$ and the strength of RSOC, $\alpha$ on the
conductance properties of a four-probe junction device. Both these
factors destroy the longitudinal $(G_L)$ and spin Hall conductances
$(G_{SH})$ for the parameter regime that we have considered in our
work, namely $0\le W\le\frac{BW}{2}$ and $0\le\alpha\le BW$ ($BW$:
bandwidth = 8$t$). For lower values of $\alpha$ ($0\le\alpha\le1$),
$G_L$ shows weak dependence on $\alpha$ and vanishes at large values
of $\alpha$. Further, $G_L$ diminishes as disorder is increased:
larger $\alpha$ registes a higher conductance at strong disorder
($W=4$).

The spin Hall conductance, $G_{SH}$ is more strongly affected by
disorder (than $G_L$) which vanishes at $W\approx1.5$ and again a
larger $\alpha$ yields larger conductance at low disorder. Further
$G_{SH}$ shows an antisymmetric behaviour for $0\le\alpha\le1$, while
it vanishes at larger $\alpha$, albeit with some fluctuations. These
fluctuations remain even after the configuration averaging being done
over 10000 disorder realizations.

Further the spin Hall conductance fluctuations, $\Delta G_{SH}$ do not
have a universal nature and shows strong dependencies on $W$ and
$\alpha$, all the while remaining at lower than its universal value
$\frac{e}{8\pi}$.

Finally we do not get any convincing evidence for a RSOC induced
transition to a metallic state and we feel that it is a physically
meaningful result as the RSOC does not break the time reversal
symmetry, a crucial condition for the weak localization to occur,
which however could have been a different scenario if an external
magnetic field would have been present.

\section{Acknowledgment} 

For most of our numerical calculations we have used KWANT
\cite{kwant}. SB thanks CSIR India for financial support under the
Grant F.No:03(1213)/12/EMR-II.

\end{document}